\documentclass[prl,twocolumn,superscriptaddress,floatfix]{revtex4}
\usepackage{graphicx}
\usepackage{times}
\usepackage{color}
\usepackage{bm}
\usepackage{array}
\usepackage[colorlinks,bookmarks=false,citecolor=blue,linkcolor=red,urlcolor=blue]{hyperref}
\usepackage{appendix}

\begin{document}

%\title{Quantum criticality of a one-dimensional Ising-anisotropic Kondo lattice model}
\title{Local Quantum Criticality of a One-Dimensional Kondo Insulator Model}
\author {W. Zhu}
\email{weizhu@lanl.gov}
\affiliation{Theoretical Division, T-4 and CNLS, Los Alamos National Laboratory, Los Alamos, New Mexico 87545, USA}
\author {Jian-Xin Zhu}
\email{jxzhu@lanl.gov}
\affiliation{Theoretical Division, T-4 and CNLS, Los Alamos National Laboratory, Los Alamos, New Mexico 87545, USA}
\affiliation{Center for Integrated Nanotechnologies, Los Alamos National Laboratory, Los Alamos, New Mexico 87545, USA}
%\author{XX}
%\affiliation{XX}

\begin{abstract}
The continuous quantum phase transition and the nature of quantum critical point (QCP) in a modified Kondo lattice model with
Ising anisotropic exchange interactions is studied within the density-matrix renormalization group algorithm.
We investigate the effect of quantum fluctuations on critical Kondo destruction QCP,
by probing static and dynamic properties of the magnetic order and the Kondo effect.
In particular, we identify that
local Kondo physics itself becomes critical at the magnetic phase transition point,
providing unbiased evidences for local quantum criticality between two insulators without resorting to the change of Fermi surface. %evidenced by the singularity of local spin susceptibility,
\end{abstract}
%\pacs{
%75.10.Jm, % quantized critility
%75.40.Mg, % numerical simulation studies
%75.40.Gb % dynamic properties (dynamic susceptibility, spin waves, spin diffusion, dynamic scaling, etc.)
%}

\date{\today}

\maketitle

%\tableofcontents

%\clearpage
\textit{Introduction.---}
Quantum criticality describes the collective fluctuations of matter undergoing a continuous phase transition at zero temperature~\cite{SSachdev2011}.
As the quantum criticality is central to a broad understanding of strongly correlated quantum matter,
how to properly describe the physics around quantum critical points (QCPs) is a subject of intensive research~\cite{Hilbert2007}.
The intermetallic heavy-fermion compounds~\cite{GStewart1984,HTsunetsugu1997,GStewart2006} serve as ideal candidates for the study of quantum phase transition and criticality, by exhibiting unusual properties like heavy-Fermi liquid, magnetic ordering, as well as unconventional superconductivity~\cite{Qimiao2008}. Recently, a continuous suppression of antiferromagnetic transition temperature has  been discovered in  a sizable number of (nearly) stoichiometric heavy fermion systems~\cite{Hilbert2007}.
%   Recently heavy-fermion systems have emerged as a playground to study QCPs. %
For QCPs relevant to the heavy-fermion systems, two major theoretical scenarios have been proposed:
One is the spin-density-wave QCP \cite{Hertz1976,Millis1993}
and the other one is critical Kondo destruction QCP \cite{Qimiao2001,Senthil2003,Coleman2010}.
For spin-density-wave QCP, conduction electrons acquire peculiar dynamics through an essentially perturbative coupling
to the slow critical modes of magnetic background.
While in the latter case,
the local Kondo physics itself becomes critical at the magnetic ordering transition,
and the QCP is driven by the competition between local dynamics and the long-ranged magnetic fluctuations.
%intense studies
Despite considerable efforts,
debate continues  on the nature of QCP,
and several issues remain elusive  in the heavy-fermion systems.
First, it is generally believed that the spin fluctuations in three dimension leads to
a Doniach's QCP~\cite{Doniach1977,Varma1976} with dynamical spin susceptibility satisfying usual Fermi-liquid form,
while two-dimensional spin fluctuations tend to favor local QCP with  spatially-extended critical degrees of freedom coexisting at the critical point \cite{Qimiao2001}. An important question is what kind of QCP the magnetic transition in one dimension could follow.
%Second, for some given model, the actual zero temperature transition in Kondo lattice model (KLM)
%is second-order or not. For example, earlier works on Ising-anisotropic KLM using various quantum Monte Carlo approaches
%led to some conflicting conclusions \cite{JXZhu2003,PSun2003}.
%explicit demonstration or identification of the
%non-Abelian states and the associated statistics in a microscopic model is very challenging.
Second, a key assumption to distinguish different scenarios usually resorts to
the shrink of Fermi surface from large to small when across a local QCP~\cite{Qimiao2008}.
Although the argument of Fermi surface in metallic phase is natural~\cite{Xavier2002,Basylko2008,Troyer1993,Ueda1993,Moukouri1996},
the QCP connecting two insulators without Fermi surface is hardly explored,
raising the question of  whether the change of Fermi surface is intrinsic to the local QCP scenario.
Numerically, the extended dynamical mean-field theory (EDMFT)~\cite{Qimiao2001,Grempel2003,JXZhu2003,JXZhu2007,PSun2003,Glossop2007a,
Glossop2007b}  and large-$N$~\cite{IPaul2007,CPepin2007} approaches have been used to
determine the nature of QCP in heavy-fermion systems. In these approaches, the spatial and temporal quantum fluctuations are either partially or completely neglected, which is valid in high dimension.
Therefore, an unbiased and accurate numerical method to capture the full quantum fluctuations of local moments and itinerant electrons, which become particularly important in low-dimensional  systems,  is highly desired to clarify the nature of QCP.

The aim of this paper is to address the aforementioned problems,
and provide compelling numerical evidences for locally critical phase transition
in a microscopic Kondo lattice model (KLM) in one dimension.
Based on the density-matrix renormalization group (DMRG) calculations,
we are able to access the low-lying energy excitations, static and dynamical correlations of local moments as well as the charge degree of freedom.
We first identify a continuous phase transition between Kondo insulator and antiferromagnetic (AFM)  phases,
signaled by the closing neutral gap and various magnetic order parameters such as magnetization.
We then demonstrate the evolution of local susceptibility across the magnetic phase transition.
The singular behavior indicates the Kondo screening being critical at the transition point, serving as the hallmark of
local quantum criticality. Importantly, we carefully clarify that the conduction electrons form spin density wave,  upon the emergence of the AFM order from the local moments. These results provide first compelling evidence of local QCP between two insulators without change of Fermi surfaces.

\begin{figure}[b]
\includegraphics[width=0.35\textwidth]{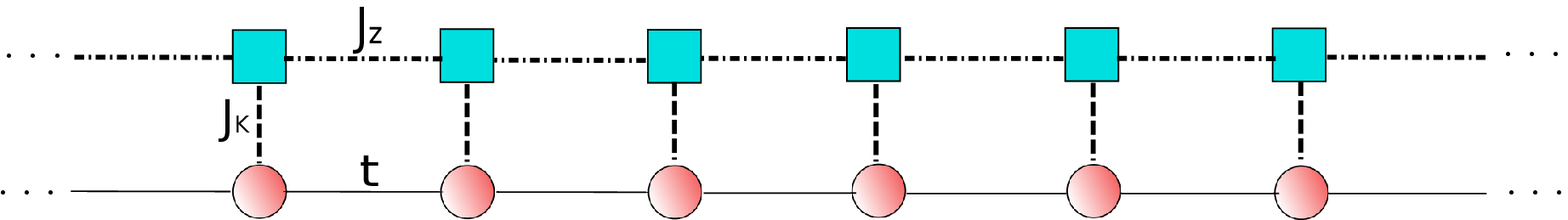}
\includegraphics[width=0.25\textwidth]{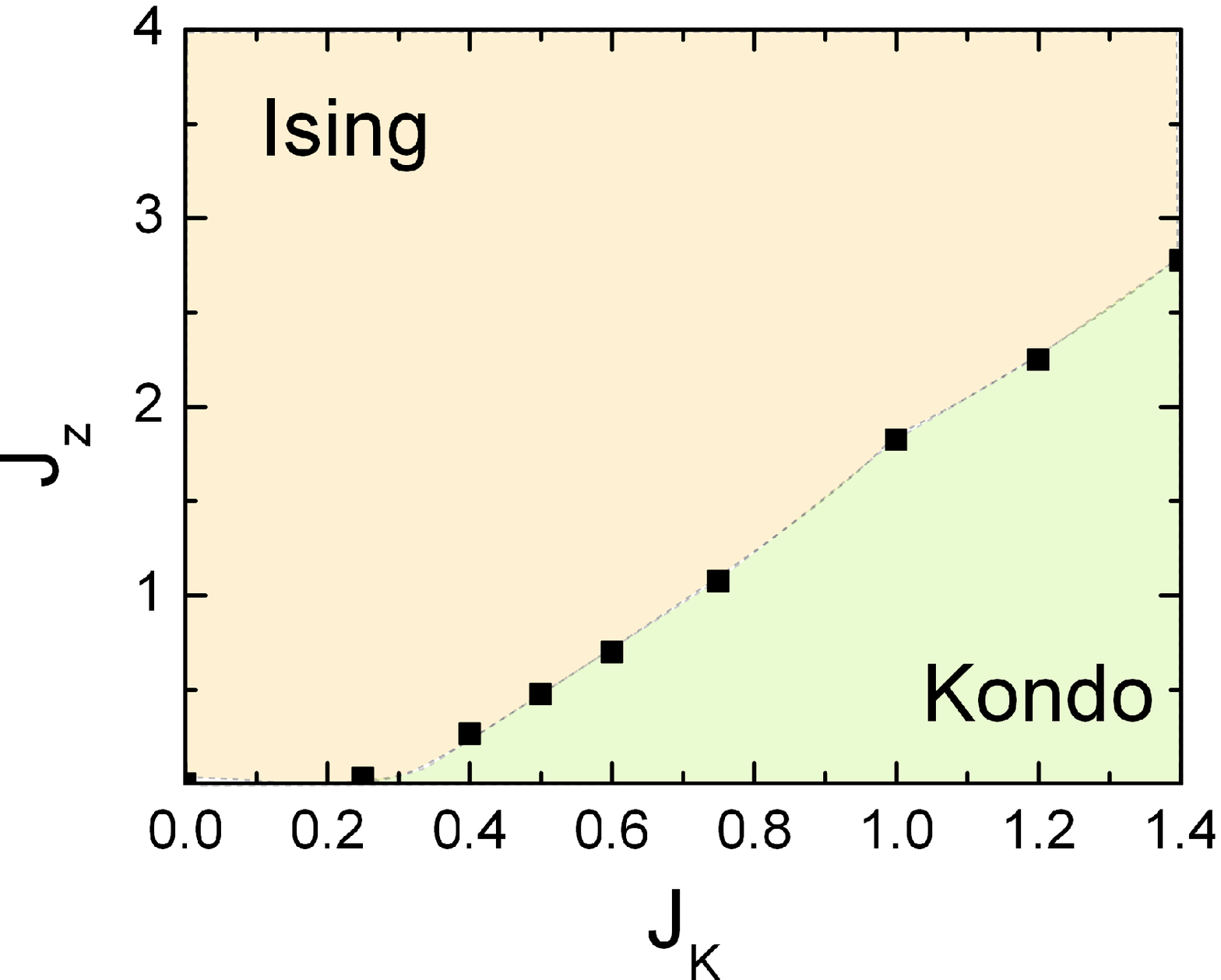}
\caption{(Top) One-dimensional Kondo lattice model with an Ising-type interaction between nearest neighbor localized spins.
Red dots and blue squares represent the conduction electrons and localized spins, respectively.
(Bottom) The global phase diagram as a function of $J_z$ and $J_K$, by setting $t=0.25$ (Bandwidth of conduction electron is $4t=1$).
The phase transition is determined to be continuous (see main text).
\label{fig:drawing}}
\end{figure}

\textit{Model and Method.---}
We consider a modified Kondo lattice model (KLM) in one dimension with an additional Ising-type interaction between the local spins (Fig. \ref{fig:drawing}),
where each unit cell contains a localized spin and an extended conduction-band electron state:
\begin{equation}\label{eq:ham}
 H=  t\sum_{\langle ij\rangle,\sigma} c^{\dagger}_{i\sigma} c_{j\sigma}  + J_K \sum_i \mathbf S_i \cdot \mathbf s_i + J_z \sum_{\langle ij\rangle}S^z_i S^z_j  \;.
 \label{eq:KLM}
\end{equation}
%{\color{blue} To Wei: I put back the factor of $\frac{1}{2}$ before the exchange interaction term. Please make sure the notation in your numerical calculation is consistent with this.}
Here $c^{\dagger}_{i\sigma}(c_{i\sigma})$ denotes the creation operator of a conduction electron  with spin $\sigma=\uparrow,\downarrow$ at site $i$.
The $\mathbf S_i$ is localized moment with $S=\frac{1}{2}$. Each localized moment interacts via an exchange coupling $J_K$ with the conduction electron,
where  the conduction electron density is defined as $\mathbf s_i=\frac{1}{2}\sum_{\sigma,\sigma'} c^\dagger_{i\alpha} \vec \sigma_{\alpha\beta}c_{i\beta}$.  The quantity $J_z$ describes the Ising-type magnetic  exchange interaction between the low moments \cite{note1}. We note that the magnetic exchange  interaction is usually generated by the Kondo interaction via the Ruderman-Kittel-Kasuya-Yosida (RKKY) effect. Here we have treated it as an independent parameter for two reasons. First, it helps the purpose of specifying the global phase diagram. Second, in one dimension, the Heisenberg-type RKKY interaction always preserves the spin-rotational invariance while the Ising interaction could stabilize AFM order~\cite{Shiba1980}. Therefore, the KLM with Ising-type exchange interaction has the advantage of ameliorating the double counting issue arising from an explicit inclusion of the intrinsic RKKY-based exchange interaction, the latter requiring a treatment of conduction electrons with care~\cite{QSi2005}.
%If a small Heisenberg term is included in KLM at half filling,
%there is no magnetic phase transition, as the ground state of one-dimensional Heisenberg model is a critical spin liquid.
Experimentally,  the easy-axis anisotropy widely exists in a number of heavy-fermion systems~\cite{HTsujii2000}.
%{\color{blue} To Wei: Can you put in a few more representative references, especially great, if we have cases for 1D heavy fermion systems.}
Physically, two important mechanisms compete with each other~\cite{Doniach1977,Varma1976}:
An isolated local moment would be screened by the spins of conduction electrons through the Kondo screening,
while the  magnetic exchange interaction tends to induce a long-ranged magnetic ordering.
%Here the Ising-type interaction between local moments plays the role to induce the magnetic order, and drive
%a magnetic phase transition from magnetically ordered phase to non-magnetic ordered phase.
In the absence of Ising-type interaction, the ground state of KLM (at half filling) is spin singlet and the spin gap always exists for any finite exchange $J_K$,
supported by both semiclassical analysis~\cite{Tsvelik1994} and finite-size numerical calculations~\cite{Tsunetsugu1992,Yu1993,Shibata1996,Shibata1999}.
In the regime where Ising-type exchange interaction dominate,  the AFM phase is expected.
Therefore, we expect a magnetic phase transition from the non-magnetic phase to the AFM phase driven by the Ising exchange interaction.

%%%%%%%%%%%%%%%%%%%%%%%%%%%%%
%%%%%%%%%%%%%%%%%%%%%%%%%%%%%

In this work, we study the KLM as described by  Eq.~(\ref{eq:KLM}) using the exact diagonalization (ED) and density-matrix renormalization group (DMRG) method \cite{White1992}.
In DMRG calculations, we use the finite system algorithm with open boundary conditions for system size up to $L=72$.
We use two different $U(1)$ quantum numbers in the DMRG set up. One is the total electron numbers $N^e=n_\uparrow+n_\downarrow$
including number of spin-$\uparrow$ $n_\uparrow$ and spin-$\downarrow$ $n_\downarrow$ electrons, the other one is the $z$-component of pseudo-spin $I^z=(n_\uparrow-n_\downarrow)/2+S^z$
where $S^z$ is the $z$-component of the total local moments.
To study the Kondo insulator we restrict ourselves to half filling where the total
number of conduction electrons $N^e$ equals number of sites $L$, or the average occupancy is one (half filling).
The dynamical response functions are computed within the scheme of dynamical DMRG~\cite{White1999,Jeck2002}.  By keeping up to $640$ states,
the truncation error is controlled below $<10^{-9}$ for static properties and $<10^{-6}$ for dynamical susceptibility calculations, respectively.

%neutral gap near I^z_c ?
%The lowest excitation is spin singlet or triplet?
%%%%%%%%%%%%%%%%%%%%%%%%%%%%%
\begin{figure}[b]
	\includegraphics[width=0.20\textwidth]{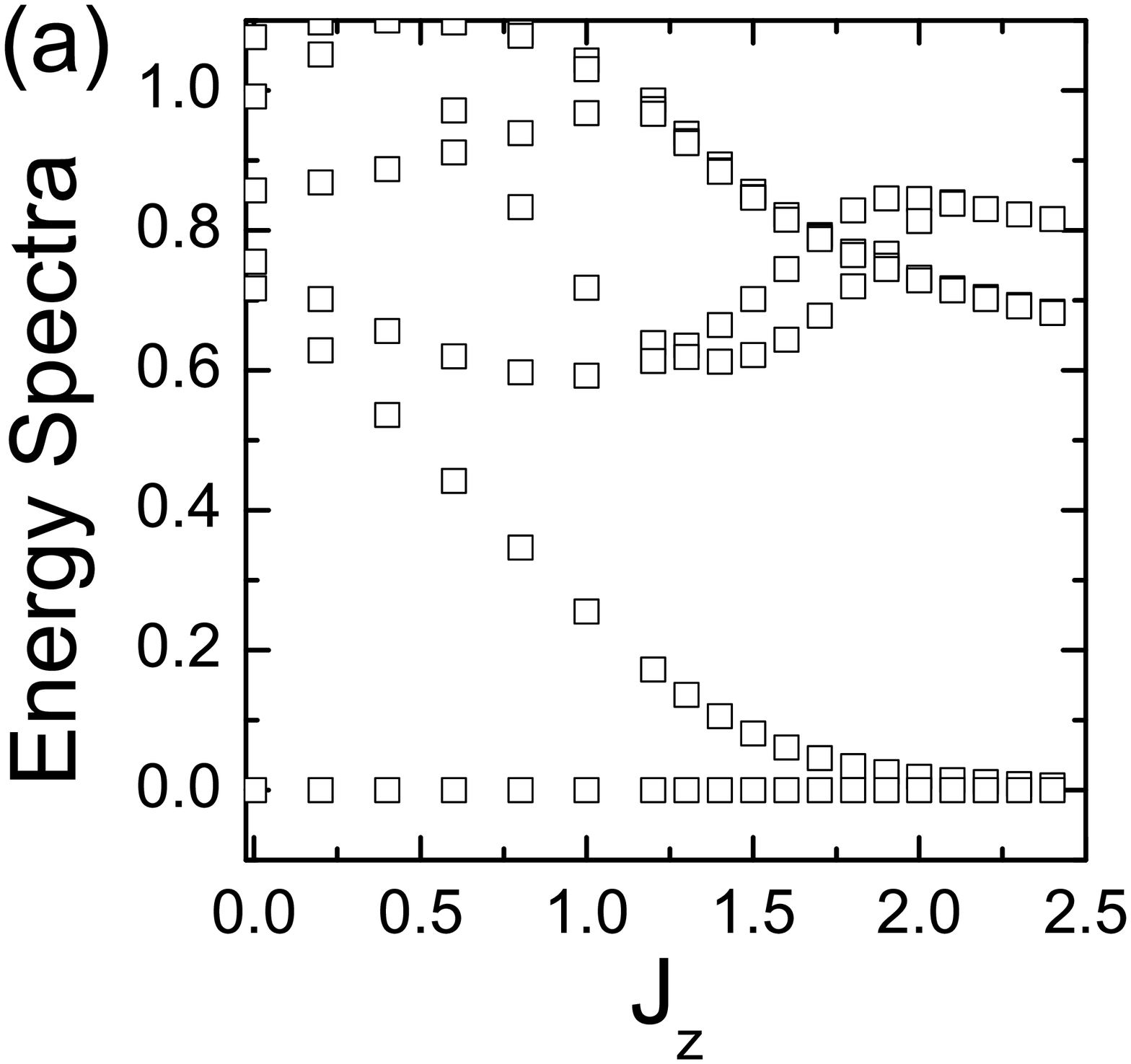}
	\includegraphics[width=0.20\textwidth]{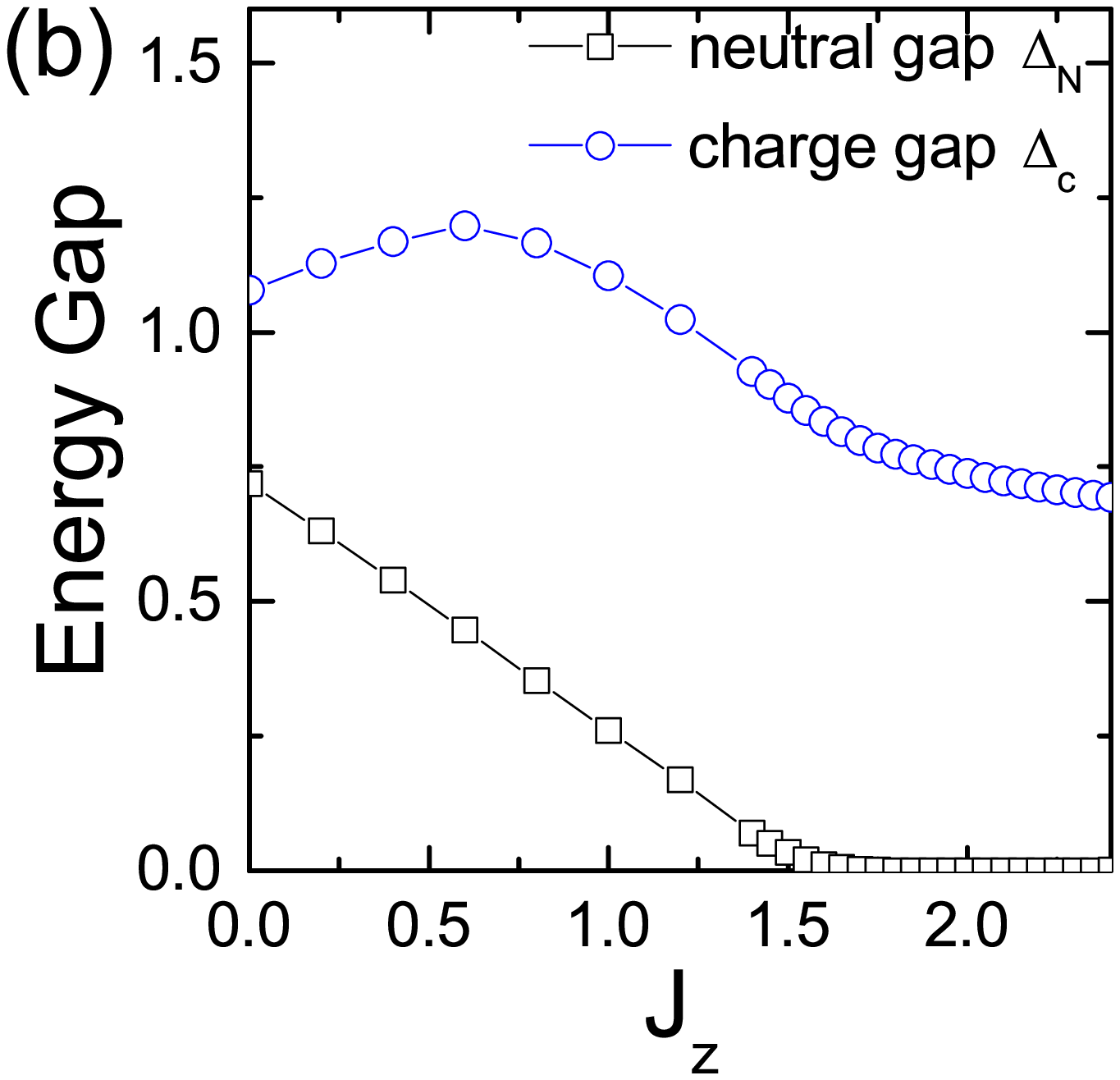}
	\caption{(a) Energy spectrum evolution as a function of $J_z$, obtained on $L=8$ periodic chain by ED calculations.
		(b) Energy gaps ($\Delta_N$, $\Delta_C$ defined in main text) as a function of $J_z$, obtained on $L=36$ open chain by DMRG calculations.
		%(c) Log-linear plot of neutral gap $\Delta_N$ near the critical point $I^z_c\approx 1.85$.
		Here we set $J_K=1.0$.
		\label{fig:gap}}
\end{figure}
%%%%%%%%%%%%%%%%%%%%%%%%%%%%%

%\clearpage
\textit{Numerical results.---}
We first present numerical evidences of Ising anisotropy driven phase transition,
based on the low-lying energy spectrum from ED calculation.
As shown in Fig~ \ref{fig:gap}(a), there exists a doublet ground state manifold in large $J_z$ regime,
relating to the AFM ground states in the Ising-limit;
while the single ground state in the small $J_z$ regime
corresponds to the ground state enclosing spin singlet between
a localized spin and one conduction electron state on each lattice site.
In particular, upon decreasing $J_z$, one energy level is continuously gapped out from the ground state manifold,
signalling a second-order type phase transition.
%Based on the ED energy spectrum, it is hard to distinguish the transition point,
Here, ED energy spectrum presents the unambiguous evidence of a continuous phase transition from
AFM ordered phase to nonmagnetic Kondo insulator phase by tuning down the Ising exchange interaction $J_z$,
whose nature will be addressed by DMRG calculations on the system of large sizes as below.

%\subsection{2. Energy gap}
Further evidences of continuous phase transition can be obtained
by DMRG calculations for larger system sizes.
Here we define two different energy gaps.
First, the energy difference
between the ground state and lowest excited state with the same quantum numbers $N^e,I^z$:
$\Delta_N=E_1(N^e=L,I^z=0)-E_0(N^e=L,I^z=0)$, is defined as the neutral gap.
%Second, spin gap is defined by the energy difference between the ground state and
%lowest excited state with different pseudo-spin number  $\Delta_S=E_0(N^e=L,I^z=1)-E_0(N^e=L,I^z=0)$.
Second, charge gap is obtained by the energy difference between ground state and lowest excited state with
different electron number $\Delta_C=E_0(N^e=L+2,I^z=0)-E_0(N^e=L,I^z=0)$.
The evolution of energy gaps as a function of $J_z$ is shown in Fig.~\ref{fig:gap}(b).
%In the absence of Ising exchange interaction, the spin gap $\Delta_S$ is equal to the neutral gap $\Delta_N$, due to the spin rotation symmetry.
By tuning up $J_z$, %the spin gap keeps open, while
the neutral gap starts to monotonically decreases to zero.
In the whole process, the charge gap is always open.
The neutral gap continuously goes to zero, supporting a second-order phase transition
driven by the  Ising anisotropy from the Kondo insulator to an Ising AFM insulating phase.

\begin{figure}
\includegraphics[width=0.2\textwidth]{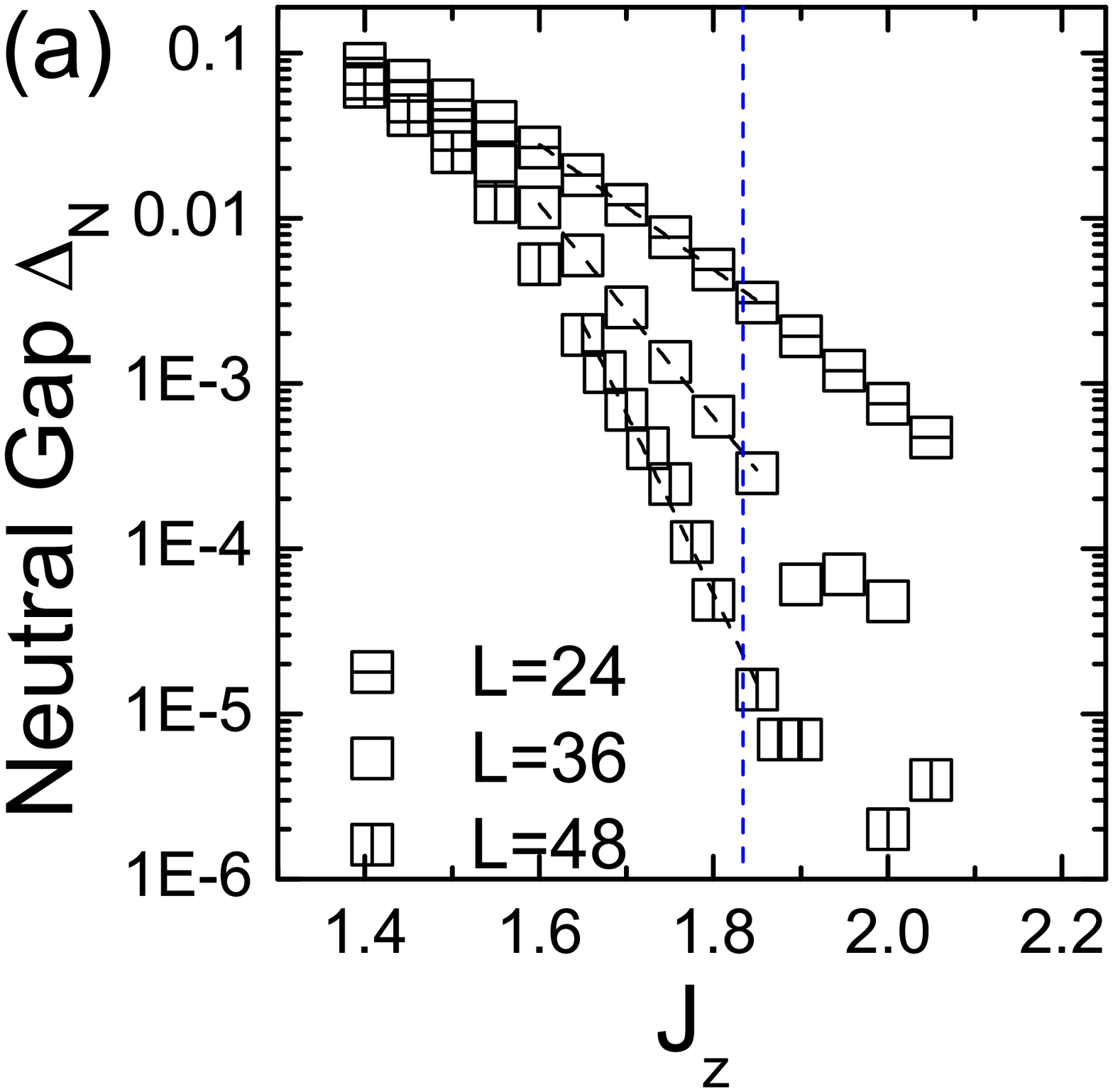}
\includegraphics[width=0.2\textwidth]{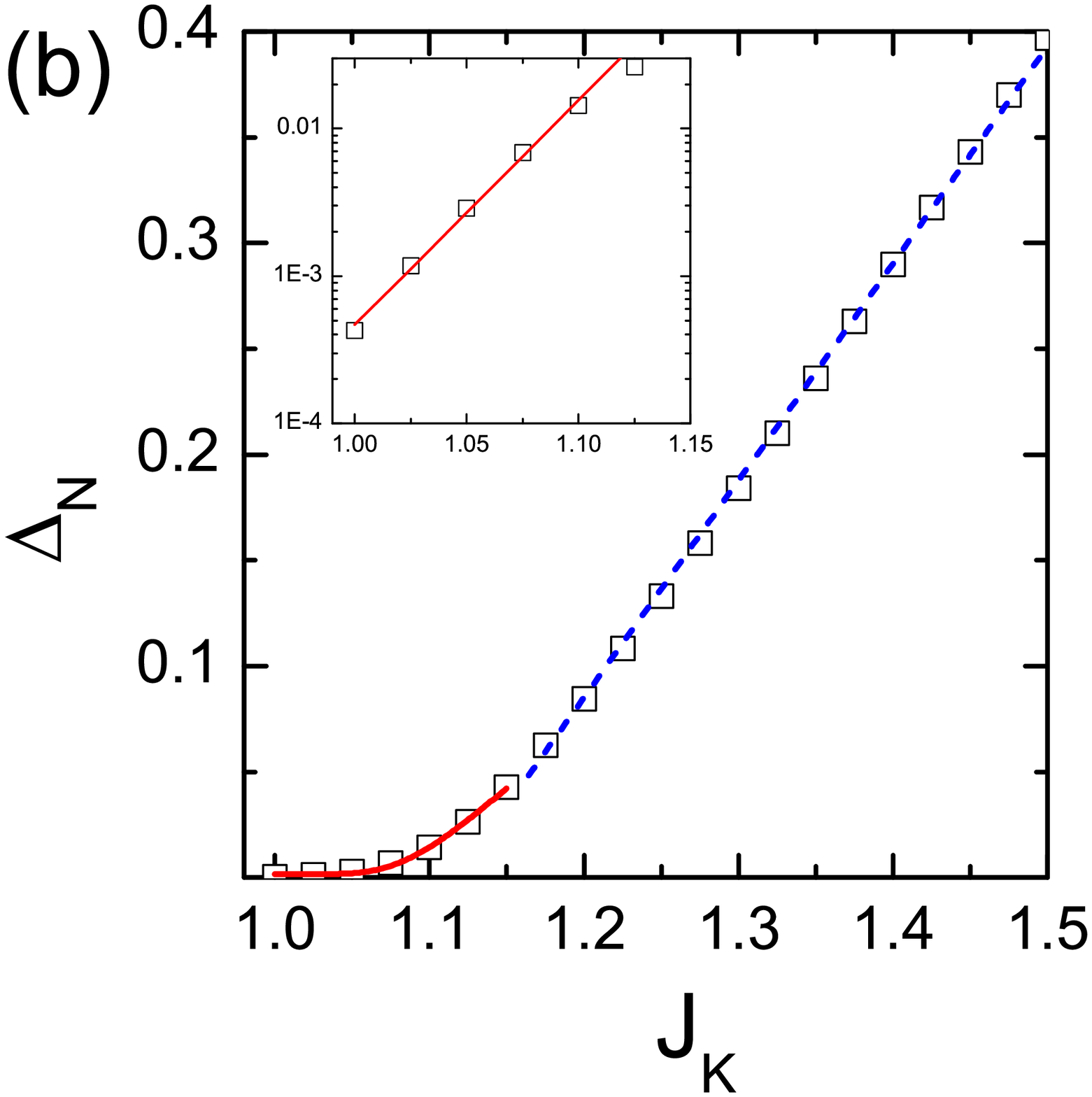}
\caption{(a) Log-linear plot of neutral gap $\Delta_N$ near the critical point $J^c_z\approx 1.825$, by setting $J_K=1.0$.
Various system sizes are labeled by different symbols.
(b) Neutral gap $\Delta_N$ as a function of $J_K$, by setting $J_z=J_z^c=1.825$.
Inset: Log-linear plot of $\Delta_N$ and exponential fitting.
\label{fig:scaling}}
\end{figure}

%\subsection{3. Scaling behavior of Energy Gap}
It is worth to mention that, the neutral gap shows exponential behavior by approaching the critical point, while away from the critical point the neutral gap is linearly dependent on $J_z$.
As shown in Fig.~\ref{fig:scaling}(a), when $J_z$
approaches the critical point, the neutral gap is found to behave as exponentially decayed,
for all system sizes.
Similarly, by tuning $J_K$, the neutral gap respectively shows  exponential dependence and linear dependence near the critical point and away from the critical point.
The exponential dependence of energy scale near the critical point is a signature of the Kondo physics becoming critical.
%: $\Delta_N\propto \exp(-\beta|I^z-I^z_c|)$ ($I^z<I^z_c$),
%where is the critical value of Ising anisotropy.

%%%%%%%%%%%%%%%%%%%%%%%%%%%%%
%%%%%%%%%%%%%%%%%%%%%%%%%%%%%

%%%%%%%%%%%%%%%%%%%%%%%%%%%%
\begin{figure}[b]
\includegraphics[width=0.4\textwidth]{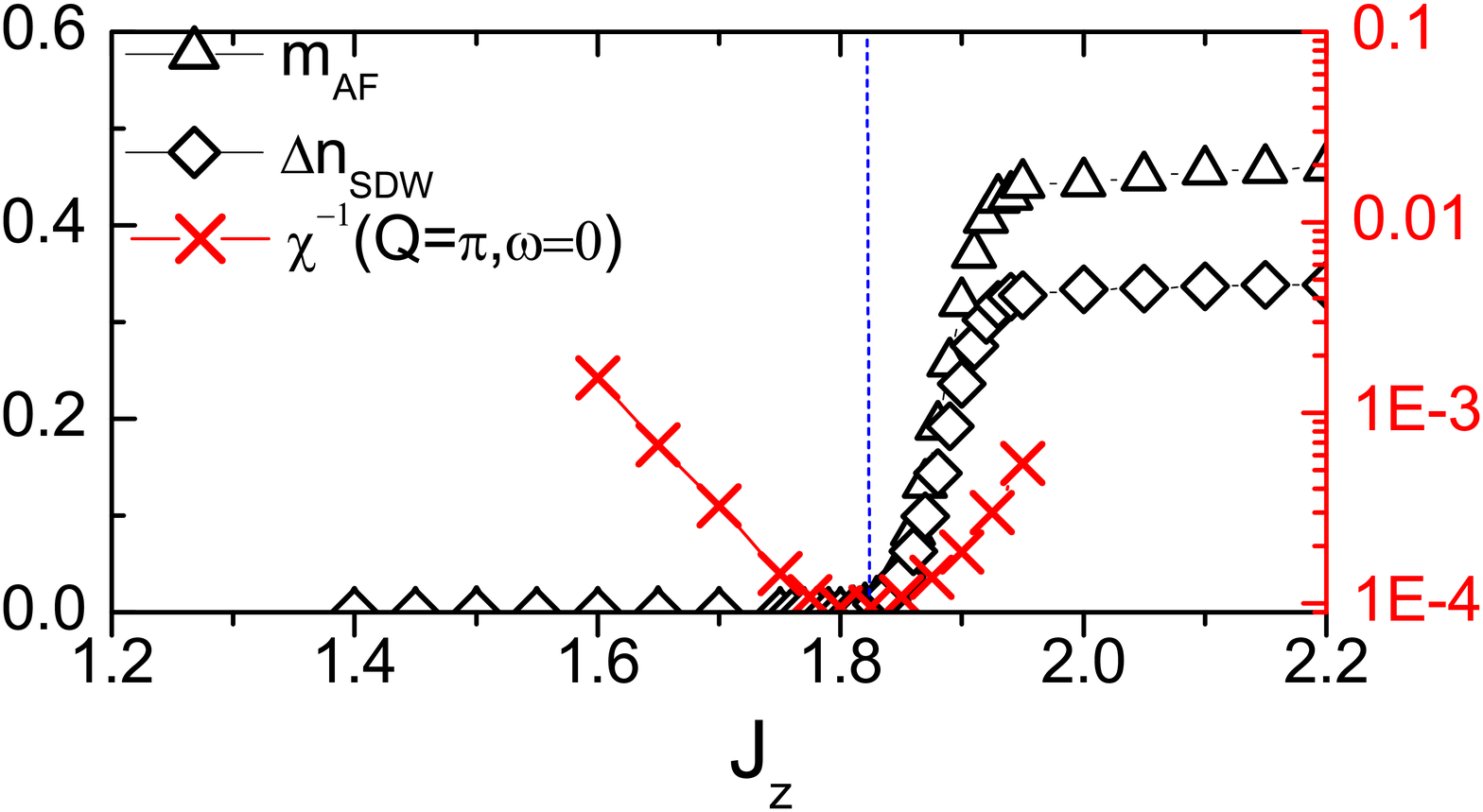}
\caption{Antiferromagnetic order parameter $m_{AF}=\frac{1}{L}\sum_i  \vert \langle S^z_i\rangle \vert$  (black triangular)
and spin density wave order parameter $\Delta n_{SDW}= \frac{1}{L} \sum_{i} \vert \langle s^z_i\rangle \vert $ (black diamond),
%{\color{blue} To Wei: Since as defined on page 2, $n_{\uparrow}$ and $n_{\downarrow}$ are the total number of spin up and down conduction electrons, I correct partially the definition of the conduction electron spin density wave. However, the site-dependent spin density wave here is still different from the definition of  $\langle s^z_i \rangle$ you define on the right column of page 3. You may want to  make this quantity consistent in the main text, figure caption, and the legend in this figure.}
and inverse static spin susceptibility $1/\chi(Q=\pi,\omega=0)$ (red cross) as a function of $J_z$.
Blue dashed line marks the transition point $J^c_z\approx 1.825$.
\label{fig:measure}}
\end{figure}
%%%%%%%%%%%%%%%%%%%%%%%%%%%%

%\subsection{4. Order Parameters}
The phase transition can be described by several local order parameters, as shown in Fig.~\ref{fig:measure}.
First, the magnetic order parameter $m_{AF}=\frac{1}{L} \sum_{i} \vert \langle S^z_i\rangle \vert$ develops continuously as $J_z$ exceeds the critical point $J_z^c$.
Importantly, we observe the charge degree of freedom shows the very similar behavior with local moments.
Within the numerical uncertainty, the spin density wave pattern ($\Delta n_{SDW}=\frac{1}{L} \sum_{i} \vert \langle s^z_i\rangle \vert $)
always occur simultaneously with nonzero magnetization $m_{AF}$.
%{\color{blue} To Wei: See my comment in the caption  of Fig. 3 on $\langle s^z_i\rangle$.}
This excludes the possibility of spin density wave driven phase transition.
In addition,
magnetic phase transition can also be probed by lattice static susceptibility at magnetic wave vector.
The lattice static susceptibility is defined as
$ \chi(Q,\omega)= -i \int dt e^{i\omega t}\langle  [S^z_{-Q}(t), S^z_Q(0)] \rangle$,
where $S^z_Q=\frac{1}{L}\sum_n \sin(\frac{\pi n}{L+1}) S^z_n$  with $n$ being the site index.
As shown in Fig.~\ref{fig:measure},
the inverse lattice static susceptibility at magnetic wave vector, $\chi^{-1}(Q=\pi,\omega=0)$ reaches a minimum at the transition point determined by $m_{AF}$ and $\Delta n_{SDW}$.
%{\color{blue} To Wei: Here when you say that the inverse susceptibility reaches a global minimum, what do you mean by the ``global minimum''? Also are you implying that it is just a minimum and does not diverge at the critical point, or you want to comment that it will be divergent when we increase the calculation resolution, e.g., reducing that broadening parameter?}
The order parameters, including lattice static susceptibility, magnetization, and charge density imbalance, point to a continuous phase transition between Kondo insulator and the AFM insulator, and
determine the magnetic critical point unambiguously.
%coincides with the critical point from $m_{AF}$ and $\langle s^z_i\rangle$
%indicates the emergence of local critical point around $J_z^c$.
%Importantly, It excludes the possibility of
%magnetic order driven by spin-density wave of charge.

%%%%%%%%%%%%%%%%%%%%%%%%%%%%%
\begin{figure}[b]
\includegraphics[width=0.45\textwidth]{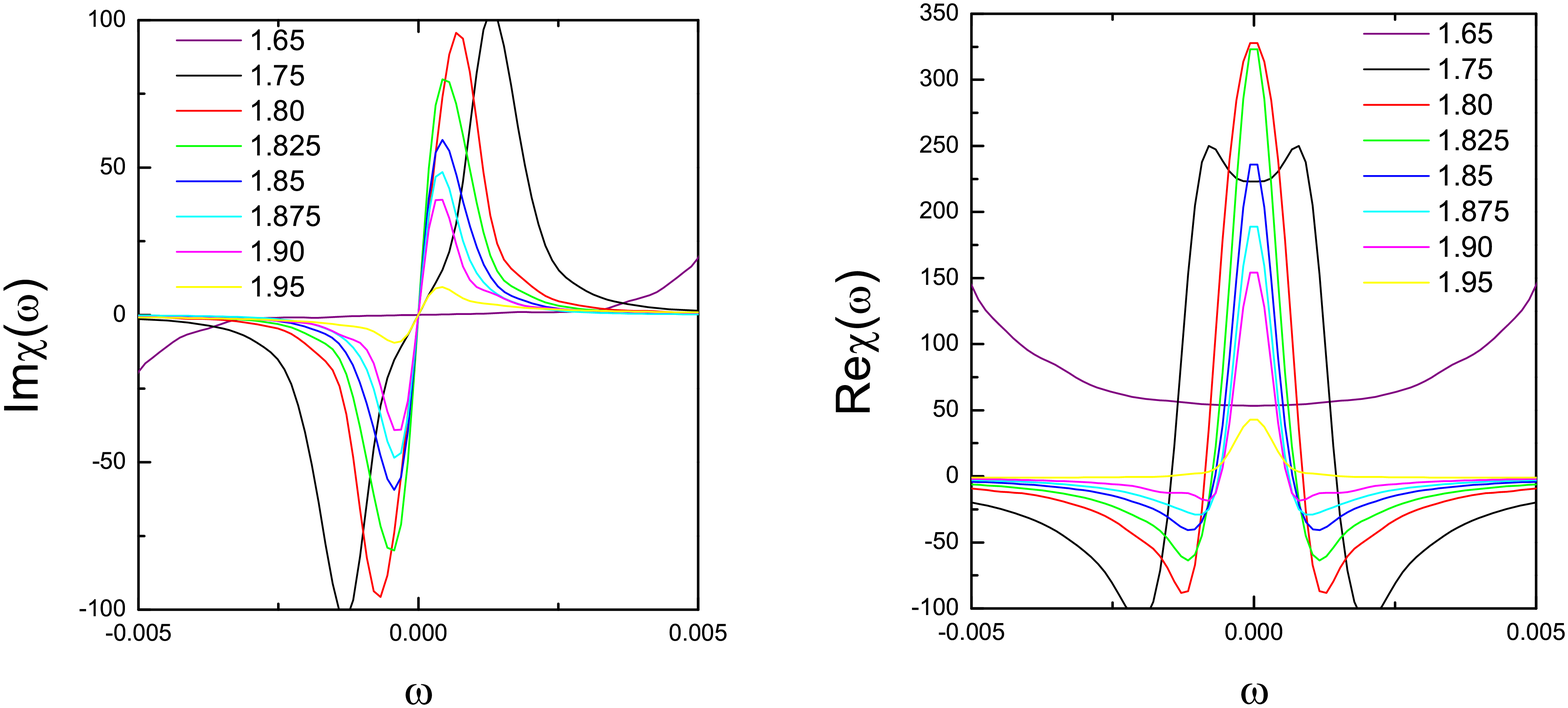}
\caption{Frequency dependence of the local spin
susceptibility at various values of $J_z$ around the magnetic transition: (Left panel) Imaginary part and (Right panel) Real part.
Here we choose $J_K=1.0$.
\label{fig:chi}}
\end{figure}
%%%%%%%%%%%%%%%%%%%%%%%%%%%%%

\begin{figure}[t]
\includegraphics[width=0.5\textwidth]{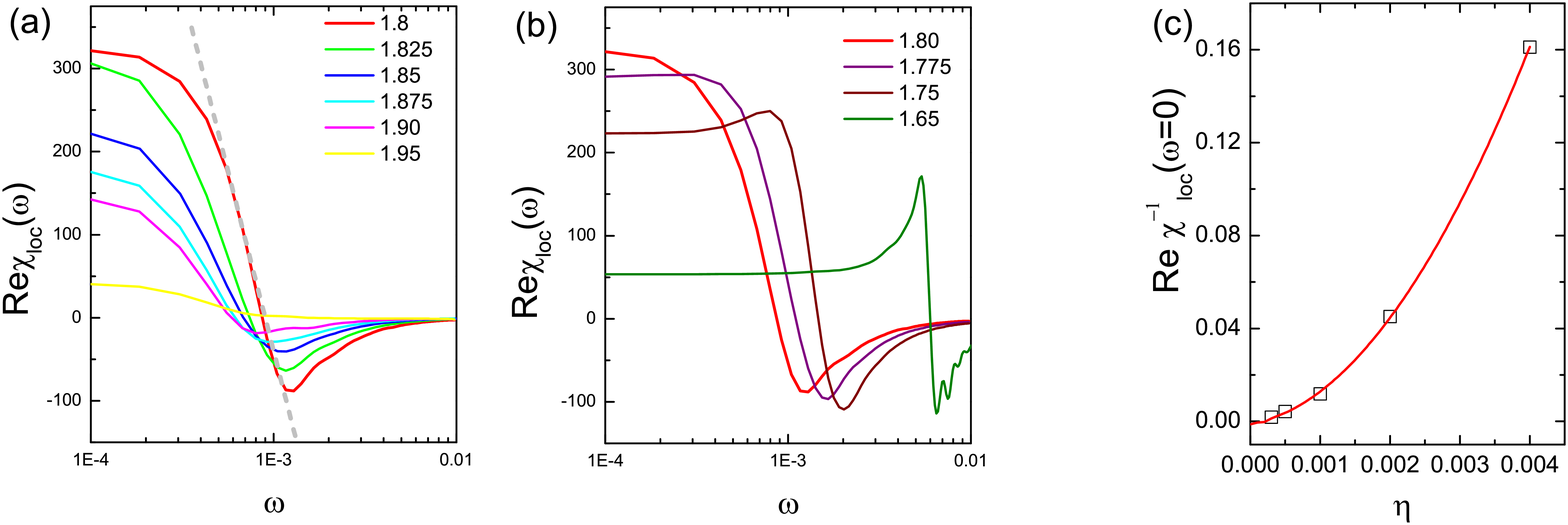}
\caption{Semi-log plot of the real part of the local spin
susceptibility around the magnetic transition: (a) $J_z>J_z^c$ and (b) $J_z<J_z^c$.
(c) Inverse of the real part of the local spin susceptibility $\Re \chi^{-1}(\omega=0)$ versus $\eta$.
Red line shows the polynomial function fitting: $\Re \chi^{-1}(\omega=0)=A \eta^2+ B\eta +C$, with nonzero $A$, $B$ and $C=-0.0013\pm0.002$.
%Inset: Log-log plot validates the power-law dependence.
\label{fig:chi_singular}}
\end{figure}

%\subsection{5. Local Susceptibility}
To uncover the nature of this phase transition, we further investigate the local dynamical response function.
For this purpose, we introduce the local spin susceptibility, which is defined as:
\begin{equation}\label{eq:chi}
 \chi_{loc}(\omega)=  \langle 0| \Delta S^z_j \frac{1}{\omega-(E_0-H)+i\eta}  \Delta S^z_j |0\rangle
\end{equation}
and $\Delta S^z_j=S^z_j - \langle S^z_j\rangle$ (we choose site $j$ in the center of the chain).
Figure~\ref{fig:chi} shows the local spin susceptibility around the quantum critical point.
In the Kondo singlet phase $J_z<J_z^c$, the peak of $\Im \chi_{loc}(\omega)$ stands away from the zero frequency.  
This peak position as a measure of the gapped spin excitations has a one-to-one correspondence to the value of the neutral gap as shown in Fig.~\ref{fig:gap}(b). As $J_z$ increases,
the dominate peak moves towards the low frequency, and reach zero frequency around $J_z\approx J^c_z$.
Near the critical point $J_z^c$, $\Im \chi_{loc}(\omega=0)$ becomes steeper, which leads to
a peak structure developing at  $\Re \chi_{loc}(\omega=0)$.
%$\Re \chi(\omega)$ develops a sharp peak at $\omega=0$.
The $J_z$ value at which $\Re \chi_{loc}(\omega=0)$ reaches the maximum defines the critical point $J^c_z$.
Since the singular behavior
$\Re \chi_{loc}(\omega)$ around zero-frequency is key to the nature of QCP,
we inspect the $\Re \chi(\omega)$ in detail in Fig. \ref{fig:chi_singular}.
We show the semi-logarithmic plot of $\Re \chi_{loc}(\omega)$ with a focus on low-frequency regime.
%The more careful analysis of the behavior $\lim_{\omega\rightarrow 0}\chi_{loc}(\omega)$ is shown in Sec. \ref{sec:decon}.
It is found that, for the Kondo insulator phase  $J_z<J_z^c$, $\Re \chi_{loc}(\omega\rightarrow 0)$ saturates to a finite value in the low-frequency limit (Fig. \ref{fig:chi_singular}(b)),
however, around the critical point $J_z\approx J_z^c$, $\Re \chi_{loc}(\omega\rightarrow 0)$ shows distinct behavior.
To demonstrate the singular behavior of $\Re \chi_{loc}(\omega=0)$, we investigate the $\Re \chi_{loc}(\omega=0)$ dependence on $\eta$,
which is imaginary part in dynamical response function Eq.~(\ref{eq:chi}).
To the best fit, we determine that the inverse of $\Re \chi_{loc}(\omega=0)$ has a polynomial dependence on $\eta$ (Fig.~\ref{fig:chi_singular}(c)).
In the intrinsic limit ($\eta\rightarrow 0$), we determine that $\Re \chi^{-1}_{loc}(\omega=0)$ is scaled to zero within the fitting accuracy,
thus $\Re \chi_{loc}(\omega=0)$ becomes singular.
Physically, the divergence of local susceptibility signals the Kondo screening being critical, which is the hallmark of local quantum
criticality~\cite{Qimiao2001,Qimiao2008}.
Here we emphasize that, compared with previous studies \cite{JXZhu2007,Glossop2007a,Glossop2007b}, the advantage of current scheme
is that we can target the behavior at zero frequency $\Re \chi_{loc}(\omega=0)$ directly, instead of relying on extracting the scaling behavior first  in the low frequency.
An additional support for critical local physics is provided by a logarithmically scaling form \cite{Qimiao2001}: $\Re \chi_{loc}(\omega) \sim \alpha \ln |\omega|^{-1}$ within energy window $T^*_K<\omega<T^0_K$, where the effective Kondo scale $T^*_K$ vanishes logarithmically slowly as approaching critical point $J_z\rightarrow J^c_z$.
In Fig. \ref{fig:chi_singular}(a),
we show such kind of scaling behavior indeed emerges in the vicinity of zero frequency (gray dashed line).

%{\color{blue} To Wei: My concern is that the referees (especially if one of them is Qimiao) may ask the question of the scaling behavior. I think we discussed on this and were short of a final answer? As long as our fitting is reasonable, it is worth putting in even if the scaling exponent is different than the previous works (there the transition is metal to metal while it is insulator to insulator here), suggesting our prediction can be tested by the inelastic neutron scattering.}

%%%%%%%%%%%%%%%%%%%%%%%%%%%%%
%%%%%%%%%%%%%%%%%%%%%%%%%%%%%

%%%%%%%%%%%%%%%%%%%%%%%%%%%%%
\begin{figure}[b]
\includegraphics[width=0.4\textwidth]{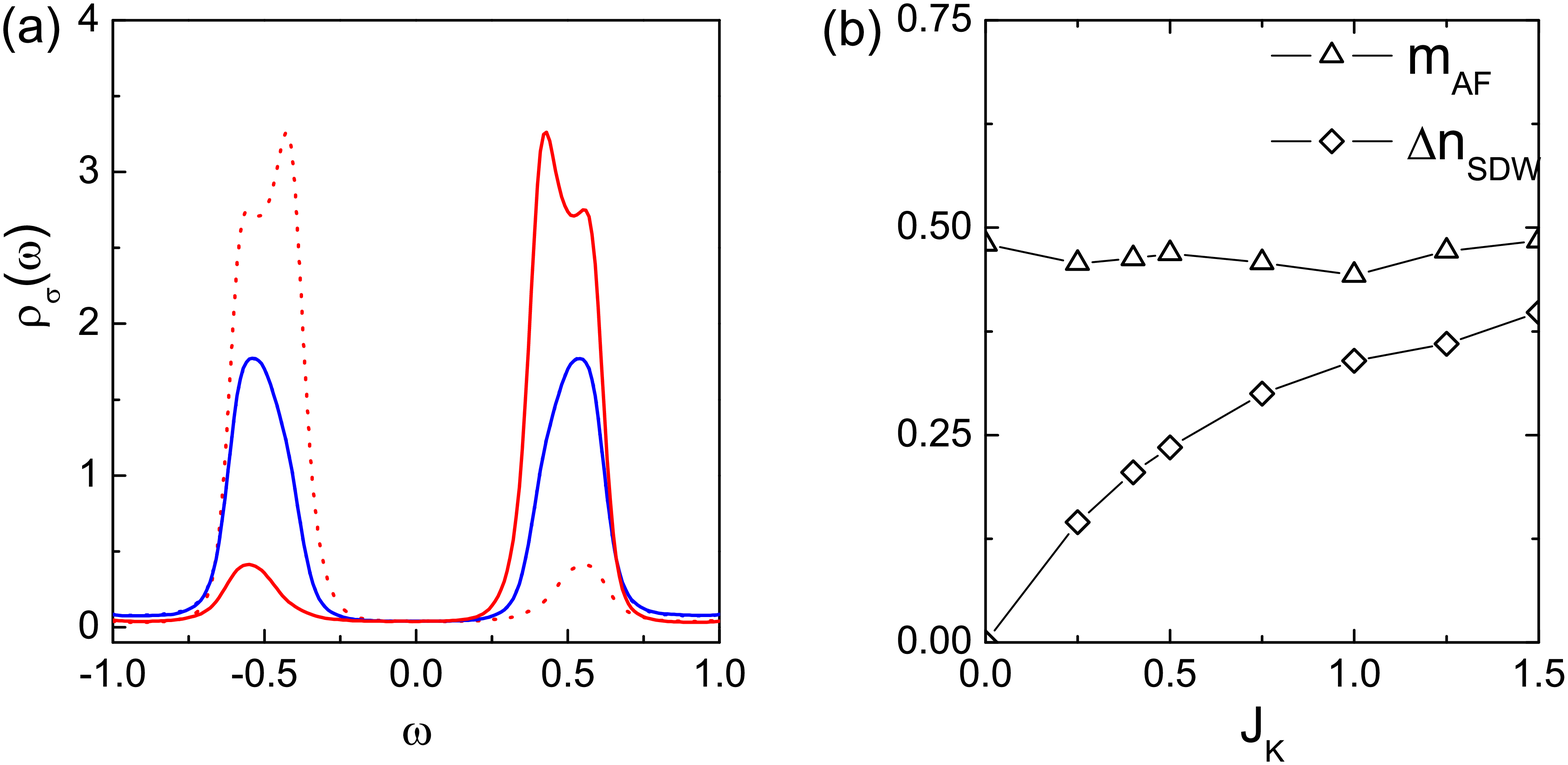}
\caption{(a) Local electron spectrum density as a function of $\omega$ for $J_z=1.65$ (blue) and $J_z=1.95$ (red).
Solid and dotted line represents spin-down and spin-up component, respectively.
%{\color{blue} To Wei: In the plot, the vertical axis label should be $\rho_{\sigma} (\omega)$, right?}
(b) In the Ising AFM phase, the $J_K$ dependence of magnetization $m_{AF}$ and charge polarization $\Delta n_{SDW}$.
%{\color{blue} To Wei: My comment on the definition of the conduction electron SDW and the legend in the plot is also applicable here.}
\label{fig:rho}}
\end{figure}
%%%%%%%%%%%%%%%%%%%%%%%%%%%%%

%\subsection{6. Charge Density}
One more advantage of our method is to treat the spin and charge degrees of freedom on an equal footing.
Here we show the electron spectrum density, $\rho_{\sigma} (\omega) = \frac{1}{L} \sum_{i}  \rho_{i\sigma}(\omega)$, around the phase transition in Fig.~\ref{fig:rho}(a), where
%\begin{equation}\label{eq:rho}
$  \rho_{i\sigma}(\omega)=-\frac{1}{\pi}\Im \langle0| c_{i\sigma} \frac{1}{\omega-(E_0-\hat H)+i\eta} c_{i\sigma}^{\dagger}|0 \rangle$.
%\end{equation}
In the Kondo insulator phase ($J_z=1.65<J^c_z$), the electron density is uniformly distributed in real space,
and the spectrum density is gapped with equal weight below and above the Fermi energy.
In the AFM phase ($J_z=1.95>J_z^c$), the spin-density wave pattern is formed in real space,
which results in an imbalance of the spectral weight of the spin-resolved spectral density in the lower and upper gap edges.
In particular, the gap around the Fermi energy in the spectrum density  remains open as $J_z$ crosses the critical point,
consistent with the charge gap evolution in Fig.~\ref{fig:gap}(b).
%This result is in striking contrast to the expectation within large-$N$ mean-field theory that the quasiparticle gap in the conduction electron sector should be closed at the critical point. {\color{blue} To Wei: Can you add more reasoning as to why?}
This result is in striking contrast to the expectation  from the Gutzwiller variational wavefunction or other auxiliary mean-field methods~\cite{JXZhu2012,JXZhu2008,Senthil2004} even for the one-dimensional systems that the quasiparticle gap in the conduction electron sector should be closed at the critical point.
In addition, we find that, in the Ising AFM phase the magnitude of spin polarization $\Delta n_{SDW}$
strongly depends on $J_K$, while the local moment magnetization $m_{AF}$ is almost unchanged.
These facts indicate that the spin-density wave in the conduction electron sector  is ``slave'' to the local spin AFM order,
partially supporting the local critical picture.
%There are two main features in charge densities. First, in the whole regime of $J_z$,
%the charge excitation gap always keeps open, as shown by a significant gap in the Fig. \ref{fig:rho}.
%Second, the charge density shows a double-peak structure.

\textit{Conclusion.---}
We have presented a thorough numerical study of a continuous phase transition between the Kondo insulator and the antiferromagnetic phases in a modified Kondo lattice model, which is of great present interest in connection with heavy-fermion quantum criticality.
Around the magnetic phase transition point, the magnetic order parameter vanishes continuously
and the static susceptibility at the magnetic ordering wave vector diverges.
A concomitant divergence of the static local susceptibility signals that the Kondo physics also becomes critical at the quantum critical point.
These results provide a ``proof-of-the-principle'' example that the local quantum criticality~\cite{Qimiao2001}  can also occur for the transition between two insulating phases,
where the Fermi surface becomes irrelevant.  It indicates that the local quantum criticality is a paradigm for novel phase transitions, which deserves to be explored in other areas of physics 
 (e.g., the interplay between strong correlation and topology in heavy fermion systems).

%{\color{blue} To Wei:  I take out your paragraph on the EDMFT and instead touch it in the paragraph below Eq.(1). Instead since we are talking about the insulating to insulating, it would be more intriguing if we can put the work into the context of  the topological phase transition and the necessity of gap closing (?) for such transitions, in particularly, when we have a Kondo break down at the magnetic transition critical point.}

%Before us, all evidences of local criticality are based on EDMFT \cite{Glossop2007a,Glossop2007b,JXZhu2003,JXZhu2007,Grempel2003}. However, the character of the quantum critical point obtained from EDMFT remains contentious. For instance, Ref. \cite{PSun2003} reported a first-order transition in periodic Anderson model, casting doubt on local criticality picture and the ability of EDMFT to capture the key physics in Kondo lattice model. Here, to our best knowledge, we provide first piece evidence beyond EDMFT. Our results from unbiased DMRG approach not only essentially yield an important confirmation
%but also shed new lights on
% local quantum criticality picture.

\textit{Acknowledgments.---}
%\acknowledgments
%WZ thanks for ... and ... for simulating discussion.
%We thank for helpful comments from the reviewers.
This work was supported by U.S. DOE at Los Alamos National
Laboratory under Contract No. DE-AC52-06NA25396 through the LANL LDRD Program (W.Z.), and U.S. DOE Office of Basic Energy Sciences (J.-X.Z.).
It was supported in part by the Center for Integrated Nanotechnologies, a U.S. DOE Basic Energy Sciences user facility.

%\clearpage

%\clearpage
%\begin{widetext}
%\widetext

%\appendix
%\begin{appendices}

%\end{appendices}

%\end{widetext}

\end{document}